\documentclass[sigconf,9pt]{acmart}

\AtBeginDocument{%
  \providecommand\BibTeX{{%
    \normalfont B\kern-0.5em{\scshape i\kern-0.25em b}\kern-0.8em\TeX}}}

\acmYear{2021}\copyrightyear{2021}
\acmConference[MAISP '21]{1st Workshop on Security and Privacy for Mobile AI}{June 24, 2021}{Virtual, WI, USA}
\acmBooktitle{1st Workshop on Security and Privacy for Mobile AI (MAISP '21), June 24, 2021, Virtual, WI, USA}
\acmPrice{15.00}
\acmDOI{10.1145/3469261.3469406}
\acmISBN{978-1-4503-8601-2/21/06}

\newcommand{\ignore}[1]{}

\newcommand{\systemName}{3D-TC2}

\usepackage{multicol}
\usepackage{multirow}
\usepackage{graphicx}
\usepackage{balance}
\usepackage{textcomp}
\usepackage{xcolor}

\begin{document}

\title{Temporal Consistency Checks to Detect LiDAR Spoofing Attacks on Autonomous Vehicle Perception}


\author{Chengzeng You}
\authornote{Both authors contributed equally.}
\affiliation{%
  \institution{Imperial College London}
  \country{United Kingdom}}
\email{chengzeng.you19@imperial.ac.uk}

\author{Zhongyuan Hau}
\authornotemark[1]
\affiliation{%
  \institution{Imperial College London}
  \country{United Kingdom}}
\email{zy.hau17@imperial.ac.uk}

\author{Soteris Demetriou}
\affiliation{%
  \institution{Imperial College London}
  \country{United Kingdom}}
\email{s.demetriou@imperial.ac.uk}

\renewcommand{\shortauthors}{C. You et al.}

\begin{abstract}
LiDAR sensors are used widely in Autonomous Vehicles for better perceiving the environment which enables safer driving decisions. Recent work has demonstrated serious LiDAR spoofing attacks with alarming consequences. In particular, model-level LiDAR spoofing attacks aim to inject fake depth measurements to elicit ghost objects that are erroneously detected by 3D Object Detectors, resulting in hazardous driving decisions. In this work, we explore the use of motion as a physical invariant of genuine objects for detecting such attacks. Based on this, we propose a general methodology, 3D Temporal Consistency Check (\systemName), which leverages spatio-temporal information from motion prediction to verify objects detected by 3D Object Detectors. Our preliminary design and implementation of a \systemName{} prototype demonstrates very promising performance, providing more than 98\% attack detection rate with a recall of 91\% for detecting spoofed Vehicle (Car) objects, and is able to achieve real-time detection at 41Hz.
\end{abstract}

\ccsdesc[500]{Security and privacy~Intrusion/anomaly detection and malware mitigation}
\ccsdesc[500]{Computing methodologies~Computer vision}
\ccsdesc[500]{Computing methodologies~Scene anomaly detection}

\keywords{Autonomous Vehicle, LiDAR, 3D Object Detection, Spoofing Attack, Attack Detection, Temporal Consistency}


\maketitle

\section{Introduction}
Light Detection and Ranging (LiDAR) units are time-of-flight measurement devices commonly used by Autonomous Vehicles (AVs) as a sensing modality critical in environment perception. LiDARs emit laser pulses near the infrared region of the electromagnetic spectrum (750nm--1.5$\mu$m). The time taken for pulses to reflect off objects in the environment are measured, generating a 3D point-cloud of the environment. Object detectors such as PointPillars\cite{lang2019pointpillars} and SECOND\cite{yan2018second} take a 3D point cloud as input and generate bounding boxes for objects in the environment, providing spatial awareness and enabling the AV to make safe navigation and control decisions.  

Unfortunately, recent work has demonstrated the feasibility of LiDAR sensor spoofing attacks and their repercussions on the visual understanding of AV's \cite{petit2015remote, shin2017illusion,cao2019adversarial, 255240, hau2021object, caoautomated}. Alarmingly, such attacks have been shown to be successful at the model-level by managing to fool 3D object detectors to erroneously output bounding boxes for non-present obstacles (ghost objects)~\cite{cao2019adversarial,255240}. These were shown to have hazardous consequences on driving decisions of end-to-end LiDAR-based perception pipelines such as the one from Baidu Apollo~\cite{cao2019adversarial}. Because increasingly more automobile manufacturers equip vehicles with such potentially vulnerable machine learning models, and since any erroneous decisions can result in accidents threatening to the safety of both the vehicle's passengers but also bystanders and other motorists, this problem becomes particularly important to address immediately. 

Indeed, in the last couple of years, there have been works that study defense strategies to mitigate such attacks on LiDAR-based perception systems. Cao et al. \cite{255240} proposed two strategies, the first is CARLO which leverage occlusion patterns to detect spoofed object. The second approach is SVF, a general machine learning architecture which takes into account physical features to obtain an object detector that is robust against such attacks. Hau et al. \cite{hau2021shadow} introduced the notion of 3D shadows for genuine objects as a physical invariant to detect spoofed objects. However, current works only leverage static information in a single frame (3D point cloud) of a target scene. 

Cao et al. \cite{caoautomated} have demonstrated that it is possible to track a moving AV and perform LiDAR point spoofing -- thus, resulting in attackers being able to perform temporal attacks. It remains unclear if the current defenses are able to defend against such temporal attacks which demonstrates a need for complementary approaches. We observe that in practice, AVs process consecutive scenes for driving decisions which carry useful spatio-temporal information. Such information can be leveraged for predicting \textit{motion} which we propose as an additional physical invariant for detecting LiDAR spoofing attacks. In the AV driving setting, we expect that objects (and their motion trajectory) should be consistent across consecutive 3D LiDAR scenes and this temporal consistency would be disturbed when an adversary introduces a fake object. Based on these observations we propose a general and modular 3-phase methodology, \emph{3D Object-detection Temporal Consistency Check} (\systemName{}) for detecting such abnormalities. Firstly, \systemName{} employs a motion prediction model that uses consecutive LiDAR point-clouds to predict the likely trajectory of objects in the scene. Then, it aligns the output of the 3D object detector on a target frame with the respective motion prediction for that scene on a common 2D representation and coordinate space. Finally, \systemName{} employs a matching strategy to detect inconsistencies. We developed a prototype implementation of an end-to-end system using the \systemName{} approach and study its effectiveness on static attacks (i.e. attack on single scene). We show that this approach is very promising in detecting LiDAR spoofing attacks demonstrating high detection rates at more than 98\% on average for spoofed \textit{Vehicle (Car)} objects.

\section{Background and Related Work}\label{sec:related}
\vspace{3pt}\noindent\textbf{LiDAR spoofing attacks.} A number of works recently proved the feasibility of LiDAR spoofing attacks~\cite{petit2015remote, shin2017illusion,cao2019adversarial, 255240, hau2021object, caoautomated}. Petit et al. \cite{petit2015remote} first introduced the LiDAR sensor attack by performing relay attacks to spoof objects further than the location of the spoofer. Shin et al. \cite{shin2017illusion} demonstrated an attack that was able to spoof up to 10 points in a 3D point-cloud at a location closer than the spoofer. Cao et al. \cite{cao2019adversarial} and Sun et al. \cite{255240} subsequently demonstrated attacks that spoofed up to 100 and 200 points in the 3D point-cloud respectively, and additionally devising strategies to use limited spoofed points to inject "ghost" objects that result in the AV making erroneous decisions. 

\vspace{3pt}\noindent\textbf{Defences against LiDAR spoofing attacks.} LiDAR spoofing attacks have safety-critical consequences that can endanger the lives of the passengers on the vehicle or even other road users such as pedestrians and cyclists. As such, there has been a few works that study defense strategies to mitigate such attacks on LiDAR-based perception systems. Cao et al. \cite{255240} proposes two strategies, the first is CARLO which leverage occlusion patterns to detect spoofed object. The second approach is SVF, a general machine learning architecture to take into account physical features to obtain an object detector that is robust against such attacks. Hau et al. \cite{hau2021shadow} introduced the notion of shadows for genuine objects and use this physical invariant to detect spoof objects. Current defences have been shown to be effective on static 3D object detection using only information from the individual target scene (i.e. attack detection on a targeted single scene of a single point-cloud), missing rich spatio-temporal information from previous frames.

\vspace{3pt}\noindent\textbf{Temporal consistency.}  Leveraging temporal consistency for attack detection has found success in other applications such as wireless sensor networks \cite{hau2019exploiting} and object detection for videos \cite{xiao2019advit}. Our work is the first to propose motion as a physical invariant for 3D objects which it leverages to perform temporal consistency checks on 3D point clouds.

\section{3D Temporal Consistency Check}\label{sec:threat_method}
We first outline the threat model we consider and then describe the methodology of 3D Object-detection Temporal Consistency Check (\systemName) and how it could be applied to detect LiDAR spoofing attacks.

\subsection{Threat Model}
We consider a LiDAR spoofing adversary that has the ability to spoof return signals of LiDAR demonstrated in \cite{petit2015remote,shin2017illusion,cao2019adversarial, 255240}. We follow closely the threat model in \cite{hau2021shadow} with the \textit{static} adversary's ($\mathcal{A}_{static}$) capabilities and goals:

\vspace{3pt}\noindent$\bullet$ \textit{Number of spoofed points.} We assume $\mathcal{A}_{static}$ enjoys state of the art sensor spoofing capabilities and can inject $\leq200$ points~\cite{255240} in a 3D scene.

\vspace{3pt}\noindent$\bullet$ \textit{Types of spoofed objects.} We consider model-level spoofing attacks able to emulate distant and occluded vehicles, pedestrians and cyclists ~\cite{cao2019adversarial, 255240, hau2021shadow}.

\vspace{3pt}\noindent$\bullet$ \textit{Knowledge.} We consider a white-box model-level spoofing adversary who has full knowledge of the internals of both the victim model and the detection mechanism.

\vspace{3pt}\noindent$\bullet$ \textit{Aims.} The adversary can launch \emph{ghost attacks} by spoofing front-near objects (5m-8m in front of the ego-vehicle).~\cite{cao2019adversarial,255240}. 

\subsection{\systemName \space Methodology}

\begin{figure*}[htbp]
    \centerline{\includegraphics[width=0.8\textwidth]{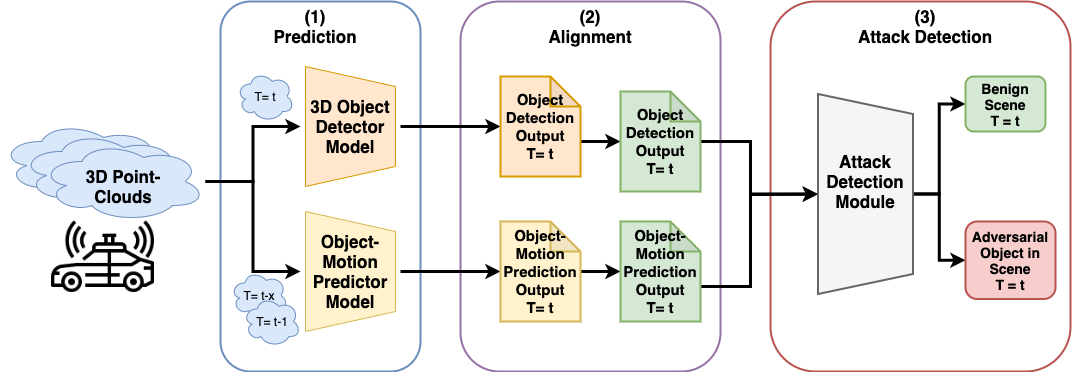}}
    \caption{Methodology of 3D-TC2. Flowchart of how 3D scenes are processed across the 3 phases and modular components.}
    \label{fig:3D-TC2_methodology}
\end{figure*}

In this work, we leverage motion as a physical invariant to verify the presence of genuine 3D objects. We propose \emph{3D Object-detection Temporal Consistency Check} (\systemName), a modular methodology which utilizes motion prediction to analyze the temporal consistency of objects across consecutive frames in a driving scene. (See Figure \ref{fig:3D-TC2_methodology}). \systemName{} consists of the following 3 Phases with a total of 4 modular components :

\vspace{3pt}\noindent\textbf{Prediction.} The prediction phase consists of 2 modular components: \textit{Object Detector} and \textit{Object-Motion Predictor}. The two components work in parallel. The \textit{Object Detector} takes in the current frame (3D point-cloud) as input and outputs object predictions as a form of bounding boxes coordinates in a 3D point-cloud. The  \textit{Object-Motion Predictor} uses historical spatio-temporal information from a number of previous 3D scenes to predict the expected location of objects in the current frame. This prediction is based on temporal information learnt from previous frames. As such, we hypothesize that any abrupt introduction of objects into a frame, which is a characteristic of LiDAR-based front-near object spoofing attack, can be detected as an anomaly. Our modular design allows for both components to be interchanged, allowing to reap the benefits from any advancements in both object detection and motion prediction. 

\vspace{3pt}\noindent\textbf{Alignment.} Although the output of the 2 components in the \textit{Prediction Phase} holds information of the current frame, the representation of these results might be different. For example, the  \textit{Object-Motion Predictor} produces object information in a 2D discretized space (i.e. predicts labels for each cell in the space) whereas the object detector provides higher-level object information in 3D space (i.e. bounding box coordinates for each object in the space). Hence, there is a need to align the prediction representations into a common representation before any useful comparison. The alignment component varies with the type of models used in the previous phase as it is dependent on the output of each model.

\vspace{2pt}\noindent\textbf{Attack Detection.} With the prediction and detection output in a common representation, we can analyze the results for any discrepancies that would indicate a potential LiDAR spoofing attack. This component is modular as well, allowing the user to interchange the strategy used for detecting discrepancies between the prediction model and object detection model for a particular frame at a specified time.

\subsection{Implementation Details}\label{sub-sec:implementation}

To demonstrate the application of our proposed methodology, we implemented a prototype based on \systemName \space for detecting ghost objects in 3D scenes. The components used in our implementation are summarized in Table \ref{tab:implementation_components}.

\begin{table}[!htbp]
\caption{Components for \systemName \space Implementation}
\label{tab:implementation_components}
\centering
\vspace{-0.3cm}
\resizebox{\linewidth}{!}{%
\begin{tabular}{llll} 
\hline
\multicolumn{1}{|c|}{\begin{tabular}[c]{@{}c@{}}\textbf{Object }\\\textbf{Detection}\\\textbf{~Model}\end{tabular}} & \multicolumn{1}{c|}{\begin{tabular}[c]{@{}c@{}}\textbf{Object-Motion }\\\textbf{Prediction }\\\textbf{Model}\end{tabular}} & \multicolumn{1}{c|}{\textbf{Alignment}} & \multicolumn{1}{c|}{\begin{tabular}[c]{@{}c@{}}\textbf{Attack }\\\textbf{Detection}\end{tabular}} \\ 
\hline
\multicolumn{1}{|l|}{\begin{tabular}[c]{@{}l@{}}PointPillars\cite{lang2019pointpillars}\\SECOND\cite{yan2018second}\end{tabular}} & \multicolumn{1}{l|}{MotionNet\cite{wu2020motionnet}} & \multicolumn{1}{l|}{\begin{tabular}[c]{@{}l@{}}Bounding Box \\Transformation\end{tabular}} & \multicolumn{1}{l|}{\begin{tabular}[c]{@{}l@{}}Cell-match Count \\Strategy\end{tabular}} \\ 
\hline
 &  &  &  \\
 &  &  & 
\end{tabular}
}
\vspace{-30pt}
\end{table}

\vspace{3pt}\noindent\textbf{Object Detection Models.} For AV perception of the current frame, we experimented with popular 3D point-cloud based object detection models such as PointPillars\cite{lang2019pointpillars} and SECOND\cite{yan2018second}. The object detection models take in the 3D point-cloud at the current frame and provide 3D bounding boxes of objects relatively to the ego-vehicle. In a benign scenario, the objects detected are all genuine, providing accurate perception of the surrounding objects to the AV. Under a LiDAR spoofing attack, $\mathcal{A}_{static}$ injects points into a scene to spoof objects. As a result, the object detection models would detect the spoofed object and this would cause the AV to make erroneous decisions (i.e. emergency brake due to a front-near object).

\vspace{3pt}\noindent\textbf{Object-Motion Prediction Model.} For prediction of the objects in the current frame from previous frames, we use a deep-learning model MotionNet\cite{wu2020motionnet}. MotionNet takes a sequence of consecutive scenes (3D point-clouds) as
input, $Time = T_{t-K}$ to $T_{t-1}$ (where $K$ is the number of historical frames and $t$ is the time for the predicted frame), and outputs a bird’s eye view (BEV) map of the predicted frame at $Time = T_{t}$. The BEV map is a 2D representation top-down view of the scene, which is then further discretized into grid cells. MotionNet predicts for each cell, the object class label and motion information. MotionNet classifies objects into Vehicles, Bike, Pedestrian, Others and Background. Where ``Vehicle'' refers to objects such as cars, buses and trucks, ``Bike`` refers to bicycles and cyclists objects, ``Others'' refers to unclassified objects not seen in training dataset and ``Background'' refers to cells with LiDAR measurements due to objects in the environment such as roads and buildings.

\vspace{3pt}\noindent\textbf{Alignment Operation.} As the output of the 3D Object Detection Models and Object-Motion Prediction Model are represented differently (3D bounding boxes vs. 2D grid cells), there is a need to align the model outputs into a common representation so that comparison of model output could be performed. In our implementation, we operated on the output of the Object Detection Models, performing transformation of the 3D bounding boxes to match the output representation of MotionNet. The bounding boxes are transformed into the same coordinate system of MotionNet output and the dimension reduced to a 2D planar representation. The 2D planar representation is then super-imposed and mapped onto the MotionNet BEV map, where the bounding boxes are aligned with the grid cells. This results in a 2D BEV map with both information from the Object Detection's output and MotionNet's output.

\begin{figure}[htbp]
    \centerline{\includegraphics[width=0.45\textwidth]{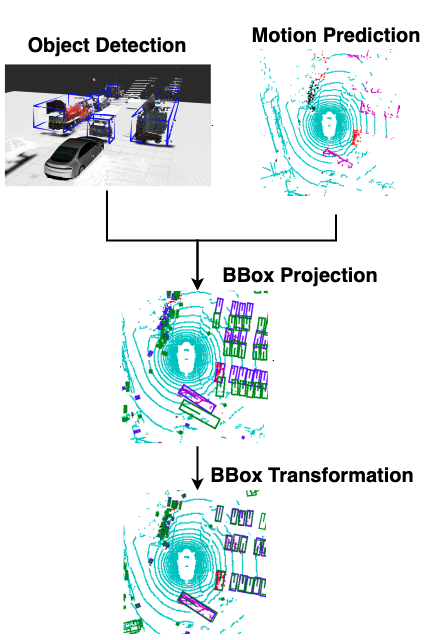}}
    \caption{Example of alignment operation.}
    \label{fig:alignment_example}
\end{figure}

Figure~\ref{fig:alignment_example} illustrates the alignment operation on a target frame, where the green boxes denote projected bounding boxes from 3D to 2D and purple boxes denotes the ground truth 2D bounding box. After the alignment operation, the projected bounding boxes matches the ground truth boxes.

\vspace{3pt}\noindent\textbf{Attack Detection Module.} This module identifies anomalous objects in a frame by comparing the aligned predicted frame  with the output of the object detection model on the actual frame. For the comparison we device and experiment with a simple \emph{Cell-match Counting Strategy} (CMCS).

\textit{CMCS} is used to determine if a detected object's bounding box matches the location predicted based on an object's motion. This is done by counting the object categories of the grid cells occupied by a detected bounding box, the object category that has the most cells in the bounding box region would be considered as the predicted object and if this object category differs from the object detection result, it is marked as a potential LiDAR spoofing attack. Under a benign scenario, the majority of the grid cell categories would correspond to the object category of the bounding box. Under a single frame LiDAR spoofing attack, when an object is successfully injected, it does not have ``history'' from the previous frames and hence there is no equivalent motion prediction of such category for the current frame. As such, the majority cell category in the bounding box of the detected spoofed object would be expected to be ``background''. We evaluate the effectiveness of CMCS in detecting single frame object spoofing attacks in Section \ref{sub-sec:attack_detect_eval}.

\section{Experiments \& Results}\label{sec:expt_res}
\vspace{3pt}\noindent\textbf{Models and Dataset.} For our evaluation, we use the models detailed in Section \ref{sub-sec:implementation}, with PointPillars\cite{lang2019pointpillars} and SECOND\cite{yan2018second} used for 3D Object Detection and MotionNet\cite{wu2020motionnet} for the Object-motion Prediction. For AV driving scenes, we use the mini dataset from nuScenes\cite{nuscenes2019}, which contains 3D point-clouds of 10 driving scenes. Each driving scene is a recording of approximately 20 sec, with a sampling rate of 2Hz. In total, each driving scene has approximately 400 frames and 40 key frames, and for the whole dataset, there are 3935 frames and 404 key frames.  

\vspace{3pt}\noindent\textbf{Evaluation of Prediction Model and Alignment.}We first evaluate the Object-motion Prediction Model choice, and the \textit{alignment} operation for transforming the output of the Object Detection Models and Object-motion Prediction Model into a common representation. We use all 3935 frames in the dataset to measure how well the prediction output of 2D BEV grid-cell map with object categories corresponds to the transformed bounding box output of the detection models. 

\vspace{3pt}\noindent\textbf{Evaluation of Attack Detection.} We evaluate the effectiveness of our implementation of \systemName \space with the \textit{CMCS} strategy in detecting model-level LiDAR spoofing attacks. For the adversarial scenarios, we consider \textbf{\textit{single-frame injection attacks}} on 362 key frames (time = $T_{key}$) in the dataset. For each key frame, a spoofed object is injected in a location of approximately 5-8m in-front of the ego-vehicle (front-near region), a threat model used in prior works \cite{cao2019adversarial, 255240} that have demonstrated spoofing objects at such distances can reliably influence driving decisions. We conduct this attack Car, Pedestrian and Cyclist objects that are commonly found in AV driving environment that have safety implications on driving decisions. These objects are also representative of objects of various sizes, allowing us to examine the effect of object sizes for our detection.

Object Detection is performed on adversarial frame at time = $T_{key}$ and MotionNet uses $T_{key - K}$\footnote{MotionNet uses K = 20 by default.} to $T_{key - 1}$ (benign) frames to predict the objects and their location at time = $T_{key}$. The results of object detection of the adversarial frame at time = $T_{key}$ and the predicted frame for time = $T_{key}$ are processed by the alignment and attack detection module to provide final detection results for objects in the adversarial frame.

\subsection{Evaluation of Prediction Model and Alignment Operation Utility}
The alignment operation is evaluated to answer the following design questions for our implementation:

\vspace{3pt}\noindent\textbf{DQ 1.} How useful is using MotionNet for the Object-motion Predictor (i.e. does object-motion prediction agrees with object detection under benign scenarios)?

\vspace{3pt}\noindent\textbf{DQ 2.} How useful is performing bounding box transformation of Object-Detection Model output (i.e. is there a good match between the output of the models under benign scenarios)?

The bounding boxes from object-detection models are transformed and overlaid onto the 2D BEV map with object categorical grid cells. The \textbf{\textit{match ratio}} for a given object category is the ratio of number of cells with that object category over the total number of cells occupied by all the bounding boxes of that object category.

\ignore{(see Eq. \ref{eq:match_ratio}).

\begin{equation}\label{eq:match_ratio}
    match\_ratio_{obj=O} = \frac{num\_cells_{obj=O}}{num\_cells\_bbox_{obj=O}}
\end{equation}
}

We further break down the analysis to objects in \textit{front-near region} and objects in \textit{front-far region}. Where front-near refers to objects in front of the ego-vehicle up to a distance of 8m from the LiDAR unit, and front-far refers to objects further than 8m. 

\vspace{3pt}\noindent\textbf{Results.} The results of our analysis are summarized in Table~\ref{tab:match_ratio} for objects in front-near and front-far regions respectively. We observed that over 80\% of the cells with the ``Vehicle'' category fall inside the ‘Vehicle’ bounding boxes demonstrating both a good prediction match and alignment. For ‘Pedestrian’ objects, over 50\% of the grid cells were found within their bounding box regions. A lower match ratio score obtained for ‘Pedestrian’ objects is due to the size of the bounding boxes which occupy much larger areas than the combined cells with predicted pedestrian categories, resulting with a large proportion of ``background'' cells. Nevertheless, our \textit{CMCS} strategy requires a majority of more than 50\% match ratio for the benign decision of the object-motion prediction agreeing with object detection. For ``Cyclist'' objects, the limited training data resulted in poor detection and prediction performance, consequently resulting in poor match. As for ‘Other’ object categories, we observed higher match in front-near regions compared to the further regions.

\vspace{3pt}\noindent\textbf{Conclusion.} From our evaluation of the match ratio between MotionNet output and the transformed bounding box output of Object Detection Models, we observe good match between the motion prediction model output and the transformed detection model output, with a match of more than 80\% for Vehicles and more than 50\% for Pedestrians at front-near and front-far regions. This indicates that the design choices of the use of MotionNet and the bounding box transformation are suitable for use in our implementation.

\begin{table*}[!htbp]
\caption{Match Ratio of Objects in Front-Near and Front-Far Regions of the Ego-Vehicle}
\label{tab:match_ratio}
\vspace{-0.3cm}
\begin{tabular}{|l|l|l|l|l|}
\hline
                    & \multicolumn{2}{c|}{\textbf{Pointpillars}} & \multicolumn{2}{c|}{\textbf{SECOND}}    \\ \hline
\textbf{\begin{tabular}[c]{@{}l@{}}Object\\ Category\end{tabular}} & \textbf{Front-Near} & \textbf{Front-Far} & \textbf{Front-Near} & \textbf{Front-Far} \\ \hline
\textbf{Vehicle(Car)}        & 9721/11500(84.53\%) & 9426/11589(81.34\%) & 9672/11192(86.42\%) & 9491/11536(82.27\%) \\ \hline
\textbf{Pedestrian} & 903/1619(55.78\%)   & 1018/1938(52.53\%)  & 826/1430(57.76\%)   & 946/1669(56.68\%)   \\ \hline
\textbf{Bike(Cyclist)}       & 12/989(1.21\%)      & 72/1112(6.47\%)     & 10/396(2.53\%)      & 78/808(9.65\%)      \\ \hline
\textbf{Other}      & 3516/5169(68.02\%)  & 3673/7644(48.05\%)  & 3428/3914(87.58\%)  & 3095/6097(50.76\%)  \\ \hline
\end{tabular}
\end{table*}

\subsection{Attack Detection Effectiveness}\label{sub-sec:attack_detect_eval}
With the prediction model and alignment module showing good utility, we are able to obtain high quality results of prediction and detection models on the same representation of a 2D BEV grid map. We now evaluate the Attack Detection Module, which uses the Cell-match Counting Strategy (CMCS) described in Section \ref{sub-sec:implementation}. This helps us answer the design question:

\vspace{3pt}\noindent\textbf{DQ 3.} How effective is the Cell-match Counting Strategy for performing a temporal consistency check to detect model-level LiDAR spoofing attacks?

For the evaluation, we generate an adversarial dataset by injecting a spoofed object into each of the 362 key frames in the original dataset at a location of approximately 8m in front of the vehicle. We define the \textbf{\textit{Attack Success Rate (ASR)}} to be the ratio of all successfully detected (by the Object Detection Model) spoofed objects over the total number of spoofed objects.

\ignore{(Eq. \ref{eq:asr}). 

\begin{equation}\label{eq:asr}
    ASR = \frac{num\_detected\_spoofed\_obj}{total\_num\_injected\_spoofed\_obj}
\end{equation}
}

Using CMCS, we are able to identify mismatches in predicted object's location with the detected object's location. Mismatched objects identified would be classified as a spoofed object. We define the metric \textbf{\textit{Detection Success Rate (DSR)}} as the ratio of all successfully identified spoofed objects (by CMCS) over the total number of successfully spoofed objects.

\ignore{(Eq. \ref{eq:dsr}). 

\begin{equation}\label{eq:dsr}
    DSR = \frac{num\_identified\_spoofed\_obj}{total\_num\_successfully\_spoofed\_obj}
\end{equation}
}

We also measure the recall of the attack detection. Recall is a metric that measures how well the detector is able to correctly identify spoofed and genuine objects.

\vspace{3pt}\noindent\textbf{Results.} Results of the effectiveness of the Attack Detection Module are summarized in Table \ref{tab:single_scene_injection_attack}. A total of 362 attack frames were used and out of these attacked frames the Attack Success Rates (ASR) for the various objects (i.e. spoofed objects detected by the victim Object Detector) are recorded and the Detection Success Rates (DSR) are successfully detected spoofed objects using temporal consistency checks. 

Our implementation of \systemName{} is able to detect spoofed ``Vehicle (Car)'' objects with an accuracy of more than 98\% and detection recall over 91\% for both detectors showing it is capable of reliably recognizing anomalous ‘Car’ objects. Detection performance for smaller objects were observed to be poorer with low detection rates for ``Pedestrian'' and ``Bike (Cyclist)'' objects. The poor performance could be attributed to MotionNet's inherent significantly poorer object classification \cite{wu2020motionnet} with classification accuracy of $\sim$77\% and $\sim$19\% for ‘Pedestrian’  and ‘Cyclist’ objects respectively. This highlights an opportunity for us to explore the use of alternative mechanisms for motion prediction in future work.

\begin{table*}[htbp]
\caption{Metrics for Single Frame Injection Attacks}
\vspace{-0.3cm}
\begin{tabular}{|l|l|l|l|l|l|l|}
\hline
\multirow{2}{*}{\textbf{\begin{tabular}[c]{@{}l@{}}Spoofed Object\\ Category\end{tabular}}} & \multicolumn{3}{c|}{\textbf{Pointpillars}} & \multicolumn{3}{c|}{\textbf{SECOND}} \\ \cline{2-7} 
                    & \textbf{ASR}     & \textbf{DSR}     & \textbf{Recall} & \textbf{ASR}     & \textbf{DSR}     & \textbf{Recall} \\ \hline
\textbf{Vehicle (Car)}    & 353/362(97.51\%) & 348/353(98.58\%) & 91.75\%         & 349/362(96.41\%) & 343/349(98.28\%) & 92.23\%         \\ \hline
\textbf{Pedestrian} & 325/362(89.78\%) & 185/325(56.92\%) & 76.93\%         & 158/362(43.65\%) & 75/158 (47.47\%) & 77.07\%         \\ \hline
\textbf{Bike (Cyclist)}    & 341/362(94.20\%) & 324/341(95.01\%) & 97.23\%         & 343/362(94.75\%) & 157/343(45.77\%) & 93.79\%         \\ \hline
\end{tabular}
\label{tab:single_scene_injection_attack}
\end{table*}

\vspace{3pt}\noindent\textbf{Conclusion.} The Cell-match Counting Strategy was able to effectively detect spoofed Vehicle objects in single-frame injection attacks at Detection Success Rate of over 98\%, as well as  recall of 91\%, for attack detection on PointPillars and SECOND. 
In all, our evaluations demonstrates that our implementation of \systemName, leveraging temporal consistency to check for valid objects in a LiDAR scene, is useful and capable of detecting spoofed object with high success rates. Compared to state-of-the-art defenses against LiDAR spoofing, our approach, with 98\% detection rate for spoofed vehicle objects, performs better than CARLO and Shadow-Catcher with detection rate of 94.5\% and 94\% respectively.

\subsection{Runtime Analysis}
Runtime constraints are important for real-time systems such as autonomous driving sensing and decision making. As such, it is important for us to be able to detect attacks in real-time in order to provide timely alerts. We perform analysis on the runtime of our implementation of \systemName \space to answer the following design question:

\vspace{3pt}\noindent\textbf{DQ 4.} How fast is the implementation of \systemName \space able to provide attack detection information?

We run our implementation of \systemName \space on the adversarial dataset 362 frames for 3 objects and measure the execution time on a machine equipped with an Intel Core i7 Six Core Processor i7-7800X (3.5GHz), 62GB RAM and 2GB NVIDIA GEFORCE GTX 1050 GPU.

\vspace{3pt}\noindent\textbf{Results.} We provide the breakdown of the runtime of the models used and the Alignment and Attack Detection components in Table \ref{tab:runtime_analysis}.  Our implementation is able to provide attack detection at approximately 41 frames per second (41Hz). From the runtime breakdown, we see that the bottleneck of the performance is in the Detection/Prediction phase, where the inference/prediction time of the models\footnotemark[2] take up the majority of the total runtime. The additional overhead introduced by our approach is approximately 5ms, which is a good trade-off in providing verification of spoofed objects. The overall runtime of $\sim$41Hz demonstrates that the implementation \systemName \space is able to provide real-time detection of spoofed objects. 

\begin{table}[htbp]
\centering
\caption{Performance / runtime of \systemName \space Components to Process A Single Frame}
\label{tab:runtime_analysis}
\vspace{-0.3cm}
\begin{tabular}{|l|l|l|} 
\hline
\multicolumn{1}{|c|}{} & \multicolumn{1}{c|}{\textbf{Mean }} & \multicolumn{1}{c|}{\textbf{std. }} \\ 
\hline
\textbf{PointPillars}\footnotemark[2] & 0.016s & - \\ 
\hline
\textbf{SECOND}\footnotemark[2] & 0.050s & - \\ 
\hline
\textbf{MotionNet}\footnotemark[2] & 0.019s & - \\ 
\hline
\textbf{Alignment} & 0.000019s (19 µs) & 0.00019s (0.19ms) \\ 
\hline
\textbf{Attack Detection} & 0.005s (5 ms) & 0.00066s (0.66ms) \\
\hline
\textbf{Total runtime} & 0.024s (24ms) &  0.85ms\\
\hline
\end{tabular}
\vspace{-5pt}
\end{table}
\footnotetext{Runtime of models are reported values from their respective papers.}

\vspace{3pt}\noindent\textbf{Conclusion.} We measured the performance of the individual components in our implementation of \systemName \space and provide the breakdown of the results. We show that the bottleneck of the detection system is attributed to the Object Detection / Prediction models, where performance can be enhanced with improvements made to state-of-the-art models. Our detection system is able to provide real-time detection at approximately 41 frames per sec (41Hz). 

\subsection{Discussion}
\vspace{3pt}\noindent\textbf{Object hiding attacks. }\systemName{} was designed to detect spoofed objects that are elicited with LiDAR spoofing attacks. Recently, there have been other classes of attacks such as Object Removal Attacks \cite{hau2021object} and MSF-ADV \cite{sp:2021:ningfei:msf-adv} that aims to hide objects from detection. We expect \systemName{} to be able to detect temporal anomalies of hiding attacks if there is abrupt disruption to the victim object. However, if the hidden object is temporally consistent (i.e. an adversarial object is placed on the road as the ego-vehicle approaches it), the approach will fail to detect such object. Detecting object hiding attacks is an interesting direction we hope to explore in future work.

\vspace{3pt}\noindent\textbf{Post-detection actions.} \systemName{} provides detection on potentially spoofed objects in a scene. As the detection is not perfect, there could be instances where it fails to detect or erroneously detect spoofed vehicles, in these instances, although rare, the AV should take a safety-first approach and prevent collision. \systemName{} can also be used in an offline fashion, as a forensic tool for post-incident analysis of 3D point-clouds for spoofed objects.



\section{Conclusion \& Future Work}\label{sec:conc_future_work}
In this paper, we proposed a general methodology, 3D Temporal Consistency Check (\systemName{}) that uses motion as a physical invariant to detect temporal inconsistencies between detected 3D objects and expected 3D objects in LiDAR-based perception systems. We implemented a prototype of \systemName \space and showed that it can detect spoofed Vehicle (Car) objects with a detection success rate of 98\% and detection recall of 91\% at 41Hz.

In future work, we intend to explore alternative approaches for motion prediction such as Kalman and particle filters, the use of alternative sequence models, and new attack detection strategies suitable for smaller objects. We also intend to consider a stronger adversary that is able to perform injection into continuous frames (temporal attacks) and study the robustness of the \systemName{} approach to such attacks.




\bibliographystyle{ACM-Reference-Format}
\bibliography{biblio}


\begin{thebibliography}{14}


\ifx \showCODEN    \undefined \def \showCODEN     #1{\unskip}     \fi
\ifx \showDOI      \undefined \def \showDOI       #1{#1}\fi
\ifx \showISBNx    \undefined \def \showISBNx     #1{\unskip}     \fi
\ifx \showISBNxiii \undefined \def \showISBNxiii  #1{\unskip}     \fi
\ifx \showISSN     \undefined \def \showISSN      #1{\unskip}     \fi
\ifx \showLCCN     \undefined \def \showLCCN      #1{\unskip}     \fi
\ifx \shownote     \undefined \def \shownote      #1{#1}          \fi
\ifx \showarticletitle \undefined \def \showarticletitle #1{#1}   \fi
\ifx \showURL      \undefined \def \showURL       {\relax}        \fi
\providecommand\bibfield[2]{#2}
\providecommand\bibinfo[2]{#2}
\providecommand\natexlab[1]{#1}
\providecommand\showeprint[2][]{arXiv:#2}

\bibitem[\protect\citeauthoryear{Caesar, Bankiti, Lang, Vora, Liong, Xu,
  Krishnan, Pan, Baldan, and Beijbom}{Caesar et~al\mbox{.}}{2019}]%
        {nuscenes2019}
\bibfield{author}{\bibinfo{person}{Holger Caesar}, \bibinfo{person}{Varun
  Bankiti}, \bibinfo{person}{Alex~H. Lang}, \bibinfo{person}{Sourabh Vora},
  \bibinfo{person}{Venice~Erin Liong}, \bibinfo{person}{Qiang Xu},
  \bibinfo{person}{Anush Krishnan}, \bibinfo{person}{Yu Pan},
  \bibinfo{person}{Giancarlo Baldan}, {and} \bibinfo{person}{Oscar Beijbom}.}
  \bibinfo{year}{2019}\natexlab{}.
\newblock \showarticletitle{nuScenes: A multimodal dataset for autonomous
  driving}.
\newblock \bibinfo{journal}{\emph{arXiv preprint arXiv:1903.11027}}
  (\bibinfo{year}{2019}).
\newblock


\bibitem[\protect\citeauthoryear{Cao, Ma, Fu, Sara, and Mao}{Cao
  et~al\mbox{.}}{2021a}]%
        {caoautomated}
\bibfield{author}{\bibinfo{person}{Yulong Cao}, \bibinfo{person}{Jiaxiang Ma},
  \bibinfo{person}{Kevin Fu}, \bibinfo{person}{Rampazzi Sara}, {and}
  \bibinfo{person}{Morley Mao}.} \bibinfo{year}{2021}\natexlab{a}.
\newblock \showarticletitle{Automated Tracking System For LiDAR Spoofing
  Attacks On Moving Targets}.
\newblock  (\bibinfo{year}{2021}).
\newblock


\bibitem[\protect\citeauthoryear{Cao, Wang, Xiao, Yang, Fang, Yang, Chen, Liu,
  and Li}{Cao et~al\mbox{.}}{2021b}]%
        {sp:2021:ningfei:msf-adv}
\bibfield{author}{\bibinfo{person}{Yulong Cao}, \bibinfo{person}{Ningfei Wang},
  \bibinfo{person}{Chaowei Xiao}, \bibinfo{person}{Dawei Yang},
  \bibinfo{person}{Jin Fang}, \bibinfo{person}{Ruigang Yang},
  \bibinfo{person}{Qi~Alfred Chen}, \bibinfo{person}{Mingyan Liu}, {and}
  \bibinfo{person}{Bo Li}.} \bibinfo{year}{2021}\natexlab{b}.
\newblock \showarticletitle{{Invisible for both Camera and LiDAR: Security of
  Multi-Sensor Fusion based Perception in Autonomous Driving Under Physical
  World Attacks}}. In \bibinfo{booktitle}{\emph{Proceedings of the 42nd IEEE
  Symposium on Security and Privacy (IEEE S\&P 2021)}}.
\newblock


\bibitem[\protect\citeauthoryear{Cao, Xiao, Cyr, Zhou, Park, Rampazzi, Chen,
  Fu, and Mao}{Cao et~al\mbox{.}}{2019}]%
        {cao2019adversarial}
\bibfield{author}{\bibinfo{person}{Yulong Cao}, \bibinfo{person}{Chaowei Xiao},
  \bibinfo{person}{Benjamin Cyr}, \bibinfo{person}{Yimeng Zhou},
  \bibinfo{person}{Won Park}, \bibinfo{person}{Sara Rampazzi},
  \bibinfo{person}{Qi~Alfred Chen}, \bibinfo{person}{Kevin Fu}, {and}
  \bibinfo{person}{Z~Morley Mao}.} \bibinfo{year}{2019}\natexlab{}.
\newblock \showarticletitle{Adversarial sensor attack on lidar-based perception
  in autonomous driving}. In \bibinfo{booktitle}{\emph{Proceedings of the 2019
  ACM SIGSAC Conference on Computer and Communications Security}}.
  \bibinfo{pages}{2267--2281}.
\newblock


\bibitem[\protect\citeauthoryear{Hau, Demetriou, Munoz-Gonz{\'a}lez, and
  Lupu}{Hau et~al\mbox{.}}{2021a}]%
        {hau2021shadow}
\bibfield{author}{\bibinfo{person}{Zhongyuan Hau}, \bibinfo{person}{Soteris
  Demetriou}, \bibinfo{person}{Luis Munoz-Gonz{\'a}lez}, {and}
  \bibinfo{person}{Emil~C Lupu}.} \bibinfo{year}{2021}\natexlab{a}.
\newblock \showarticletitle{Shadow-catcher: Looking into shadows to detect
  ghost objects in autonomous vehicle 3d sensing}.
\newblock \bibinfo{journal}{\emph{arXiv preprint arXiv:2008.12008}}
  (\bibinfo{year}{2021}).
\newblock


\bibitem[\protect\citeauthoryear{Hau, Kenneth, Demetriou, and Lupu}{Hau
  et~al\mbox{.}}{2021b}]%
        {hau2021object}
\bibfield{author}{\bibinfo{person}{Zhongyuan Hau}, \bibinfo{person}{T Kenneth},
  \bibinfo{person}{Soteris Demetriou}, {and} \bibinfo{person}{Emil~C Lupu}.}
  \bibinfo{year}{2021}\natexlab{b}.
\newblock \showarticletitle{Object Removal Attacks on LiDAR-based 3D Object
  Detectors}. In \bibinfo{booktitle}{\emph{Workshop on Automotive and
  Autonomous Vehicle Security (AutoSec)}}, Vol.~\bibinfo{volume}{2021}.
  \bibinfo{pages}{25}.
\newblock


\bibitem[\protect\citeauthoryear{Hau and Lupu}{Hau and Lupu}{2019}]%
        {hau2019exploiting}
\bibfield{author}{\bibinfo{person}{Zhongyuan Hau} {and} \bibinfo{person}{Emil~C
  Lupu}.} \bibinfo{year}{2019}\natexlab{}.
\newblock \showarticletitle{Exploiting correlations to detect false data
  injections in low-density wireless sensor networks}. In
  \bibinfo{booktitle}{\emph{Proceedings of the 5th on Cyber-Physical System
  Security Workshop}}. \bibinfo{pages}{1--12}.
\newblock


\bibitem[\protect\citeauthoryear{Lang, Vora, Caesar, Zhou, Yang, and
  Beijbom}{Lang et~al\mbox{.}}{2019}]%
        {lang2019pointpillars}
\bibfield{author}{\bibinfo{person}{Alex~H Lang}, \bibinfo{person}{Sourabh
  Vora}, \bibinfo{person}{Holger Caesar}, \bibinfo{person}{Lubing Zhou},
  \bibinfo{person}{Jiong Yang}, {and} \bibinfo{person}{Oscar Beijbom}.}
  \bibinfo{year}{2019}\natexlab{}.
\newblock \showarticletitle{Pointpillars: Fast encoders for object detection
  from point clouds}. In \bibinfo{booktitle}{\emph{Proceedings of the IEEE/CVF
  Conference on Computer Vision and Pattern Recognition}}.
  \bibinfo{pages}{12697--12705}.
\newblock


\bibitem[\protect\citeauthoryear{Petit, Stottelaar, Feiri, and Kargl}{Petit
  et~al\mbox{.}}{2015}]%
        {petit2015remote}
\bibfield{author}{\bibinfo{person}{Jonathan Petit}, \bibinfo{person}{Bas
  Stottelaar}, \bibinfo{person}{Michael Feiri}, {and} \bibinfo{person}{Frank
  Kargl}.} \bibinfo{year}{2015}\natexlab{}.
\newblock \showarticletitle{Remote attacks on automated vehicles sensors:
  Experiments on camera and lidar}.
\newblock \bibinfo{journal}{\emph{Black Hat Europe}}  \bibinfo{volume}{11}
  (\bibinfo{year}{2015}), \bibinfo{pages}{2015}.
\newblock


\bibitem[\protect\citeauthoryear{Shin, Kim, Kwon, and Kim}{Shin
  et~al\mbox{.}}{2017}]%
        {shin2017illusion}
\bibfield{author}{\bibinfo{person}{Hocheol Shin}, \bibinfo{person}{Dohyun Kim},
  \bibinfo{person}{Yujin Kwon}, {and} \bibinfo{person}{Yongdae Kim}.}
  \bibinfo{year}{2017}\natexlab{}.
\newblock \showarticletitle{Illusion and dazzle: Adversarial optical channel
  exploits against lidars for automotive applications}. In
  \bibinfo{booktitle}{\emph{International Conference on Cryptographic Hardware
  and Embedded Systems}}. Springer, \bibinfo{pages}{445--467}.
\newblock


\bibitem[\protect\citeauthoryear{Sun, Cao, Chen, and Mao}{Sun
  et~al\mbox{.}}{2020}]%
        {255240}
\bibfield{author}{\bibinfo{person}{Jiachen Sun}, \bibinfo{person}{Yulong Cao},
  \bibinfo{person}{Qi~Alfred Chen}, {and} \bibinfo{person}{Z.~Morley Mao}.}
  \bibinfo{year}{2020}\natexlab{}.
\newblock \showarticletitle{Towards Robust LiDAR-based Perception in Autonomous
  Driving: General Black-box Adversarial Sensor Attack and Countermeasures}. In
  \bibinfo{booktitle}{\emph{29th {USENIX} Security Symposium ({USENIX} Security
  20)}}. \bibinfo{publisher}{{USENIX} Association}, \bibinfo{pages}{877--894}.
\newblock
\showISBNx{978-1-939133-17-5}
\urldef\tempurl%
\url{https://www.usenix.org/conference/usenixsecurity20/presentation/sun}
\showURL{%
\tempurl}


\bibitem[\protect\citeauthoryear{Wu, Chen, and Metaxas}{Wu
  et~al\mbox{.}}{2020}]%
        {wu2020motionnet}
\bibfield{author}{\bibinfo{person}{Pengxiang Wu}, \bibinfo{person}{Siheng
  Chen}, {and} \bibinfo{person}{Dimitris~N Metaxas}.}
  \bibinfo{year}{2020}\natexlab{}.
\newblock \showarticletitle{MotionNet: Joint Perception and Motion Prediction
  for Autonomous Driving Based on Bird's Eye View Maps}. In
  \bibinfo{booktitle}{\emph{Proceedings of the IEEE/CVF Conference on Computer
  Vision and Pattern Recognition}}. \bibinfo{pages}{11385--11395}.
\newblock


\bibitem[\protect\citeauthoryear{Xiao, Deng, Li, Lee, Edwards, Yi, Song, Liu,
  and Molloy}{Xiao et~al\mbox{.}}{2019}]%
        {xiao2019advit}
\bibfield{author}{\bibinfo{person}{Chaowei Xiao}, \bibinfo{person}{Ruizhi
  Deng}, \bibinfo{person}{Bo Li}, \bibinfo{person}{Taesung Lee},
  \bibinfo{person}{Benjamin Edwards}, \bibinfo{person}{Jinfeng Yi},
  \bibinfo{person}{Dawn Song}, \bibinfo{person}{Mingyan Liu}, {and}
  \bibinfo{person}{Ian Molloy}.} \bibinfo{year}{2019}\natexlab{}.
\newblock \showarticletitle{Advit: Adversarial frames identifier based on
  temporal consistency in videos}. In \bibinfo{booktitle}{\emph{Proceedings of
  the IEEE/CVF International Conference on Computer Vision}}.
  \bibinfo{pages}{3968--3977}.
\newblock


\bibitem[\protect\citeauthoryear{Yan, Mao, and Li}{Yan et~al\mbox{.}}{2018}]%
        {yan2018second}
\bibfield{author}{\bibinfo{person}{Yan Yan}, \bibinfo{person}{Yuxing Mao},
  {and} \bibinfo{person}{Bo Li}.} \bibinfo{year}{2018}\natexlab{}.
\newblock \showarticletitle{Second: Sparsely embedded convolutional detection}.
\newblock \bibinfo{journal}{\emph{Sensors}} \bibinfo{volume}{18},
  \bibinfo{number}{10} (\bibinfo{year}{2018}), \bibinfo{pages}{3337}.
\newblock


\end{thebibliography}


\end{document}